\documentclass[%
 aip,
 amsmath,amssymb,
 reprint,%
]{revtex4-1}

\usepackage{graphicx}% Include figure files
\usepackage{dcolumn}% Align table columns on decimal point
\usepackage{bm}% bold math
\usepackage[utf8]{inputenc}
\usepackage[T1]{fontenc}
\usepackage{mathptmx}
\usepackage{etoolbox}
\usepackage{comment}
\usepackage{color}

\makeatletter
\def\@email#1#2{%
 \endgroup
 \patchcmd{\titleblock@produce}
  {\frontmatter@RRAPformat}
  {\frontmatter@RRAPformat{\produce@RRAP{*#1\href{mailto:#2}{#2}}}\frontmatter@RRAPformat}
  {}{}
}%
\makeatother
\begin{document}

\preprint{AIP/123-QED}

\title{Compact nondestructive generator for microsecond-class ultrafast pulsed magnetic fields}
% Force line breaks with \\
\author{Y. Ishii}
 \email{yutoishii@issp.u-tokyo.ac.jp}
 \affiliation{Institute for Solid State Physics, University of Tokyo, Kashiwa, Chiba 277-8581, Japan}%Lines break automatically or can be forced with \\
 
\author{A. Ikeda}%
\affiliation{Department of Engineering Science, University of Electro-Communications, Chofu, Tokyo 182-8585, Japan%\\This line break forced with \textbackslash\textbackslash
}%
\author{H. Sun}%
\affiliation{Institute for Solid State Physics, University of Tokyo, Kashiwa, Chiba 277-8581, Japan%\\This line break forced with \textbackslash\textbackslash
}%
\author{H. Oike}%
\affiliation{Research Center for Materials Nanoarchitechtonics, National Institute for Material Science, Tsukuba, Ibaraki 305-0044, Japan%\\This line break forced with \textbackslash\textbackslash
}%
\author{Y. H. Matsuda}
\affiliation{Institute for Solid State Physics, University of Tokyo, Kashiwa, Chiba 277-8581, Japan%\\This line break forced% with \\
}%

\date{\today}% It is always \today, today,

\begin{abstract}
The sweep rate of pulsed magnetic fields is a key factor in exploring of nonequilibrium phenomena in matter.
However, the nondestructive generation of ultrafast pulsed magnetic fields is challenging because of the large time constants associated with the large inductance and capacitance of conventional pulsed magnetic field generators, as well as the residual inductance of the power devices and the system itself.
We have developed a new compact pulsed magnetic field generator using commercially available state-of-the-art ultrafast power devices, enabling the generation of ultrafast pulsed magnetic fields up to approximately 1 T with a pulse duration of 3.8 $\mu$s.
We present the performance of the newly developed system and compare the experimentally obtained magnetic field waveforms with the results of a simple circuit simulation.
In addition, we demonstrate a magnetization measurement of cobalt powder and a Faraday rotation measurement of CdS single crystal using the developed system.
\end{abstract}

\maketitle

\section{\label{sec:level1}Introduction}

Magnetic field is one of the most important experimental parameters for investigating the physical properties of condensed matter systems.
Responses to magnetic fields, such as magnetic susceptibility, magnetoresistance, and magnetostriction, provide the valuable information about materials.
Such measurements allow us to elucidate the physical properties and mechanism underlying phase transitions.
High magnetic fields enable us to explore the new electronic state and phase transition.
High magnetic fields up to 45 T can be generated by a superconductor-based steady magnetic field generator enabling us to measure physical properties with very high precision \cite{SHahn2019}.
In pulsed high magnetic fields, an 80 T class field can be generated and it is essential to investigate the entire magnetization of material and the Fermi surface of material through the measurement of quantum oscillation.
Ultrahigh magnetic fields above 100 T are only accessible in a destructive manner \cite{FHerlach1973, OPortugall1997,NMiura2001,DNakamura2018}.
The ultrahigh magnetic fields can give nonperturbative effects on matters, promising way to explore the novel electronic and crystal structure \cite{TNomura2014,YHMatsuda2020,AIkeda2023,NMatsuyama2026}.

The magnetic field sweep rate is another important parameter.
The energy scales of magnetic interaction ($\sim$ 10 - 100 K) and the Zeeman splitting induced by an external magnetic field ($B$ = 1 - 100 T, corresponding to $\sim$ 1 - 135 K) typically correspond to characteristic fluctuation timescales ($\sim \hbar/k_{\rm B}T$) ranging from tens to hundreds of femto-secondes.
However, the relaxation timescale associated with a magnetic phase transition can be much longer than this simple estimate \cite{HOike2016,KMatsuura2021,KMatsuura2023}.
In such cases, thermalization can be slower than the magnetic field sweep rate, leading to the emergence of nonequilibrium state \cite{YHMatsuda2022,YIshii2023}.
Indeed, fast magnetic field sweeps provide access to nonequilibrium phenomena, such as spin-crossover associated with lattice relaxation \cite{SKimura2005}, quantum tunneling in cluster magnets \cite{RGrossinger2001,ASMishchenko2003}, and meatastable magnetization processes \cite{YHMatsuda2022}.
Such slow thermalization dynamics can arise from spin-lattice coupling and many-body effects in spin systems.
Therefore, faster magnetic field sweeps are expected to provide a powerful means of exploring nonequilibrium states in systems with multiple degrees of freedom, strong coupling between these degrees of freedom, and pronounced many-body effects.

However, the pulse duration of the nondestructive pulsed high field is typically limited to the millisecond order because of the residual inductance coming from mainly capacitor banks and switches.
As a result, the microsecond-class ultrafast magnetic fields have so far been available only using destructive techniques.
In this work, we have successfully developed the compact nondestructive generator capable of producing microsecond-class ultrafast magnetic fields up to 1 T by utilizing the state-of-the-art ultrafast powder devices.
We report the design and performance of the newly developed system.
The magnetic field waveform can be reproduced by the simple simulation that takes into account the residual inductance and resistance of the connecting cables.
Furthermore, the result of magnetization measurement of cobalt powder and the Faraday rotation on CdS single crystal will be also presented as demonstrations of applications for condensed matter physics.
Owing to the use of compact ultrafast power devices, the capacitor bank system is only approximately 100 mm in size, we refer to this new system as the "petit-bank system".
Finally, we will discuss possible further developments and applications of the petit-bank system.

\begin{figure}[h]
\includegraphics[clip,width=0.9\columnwidth]{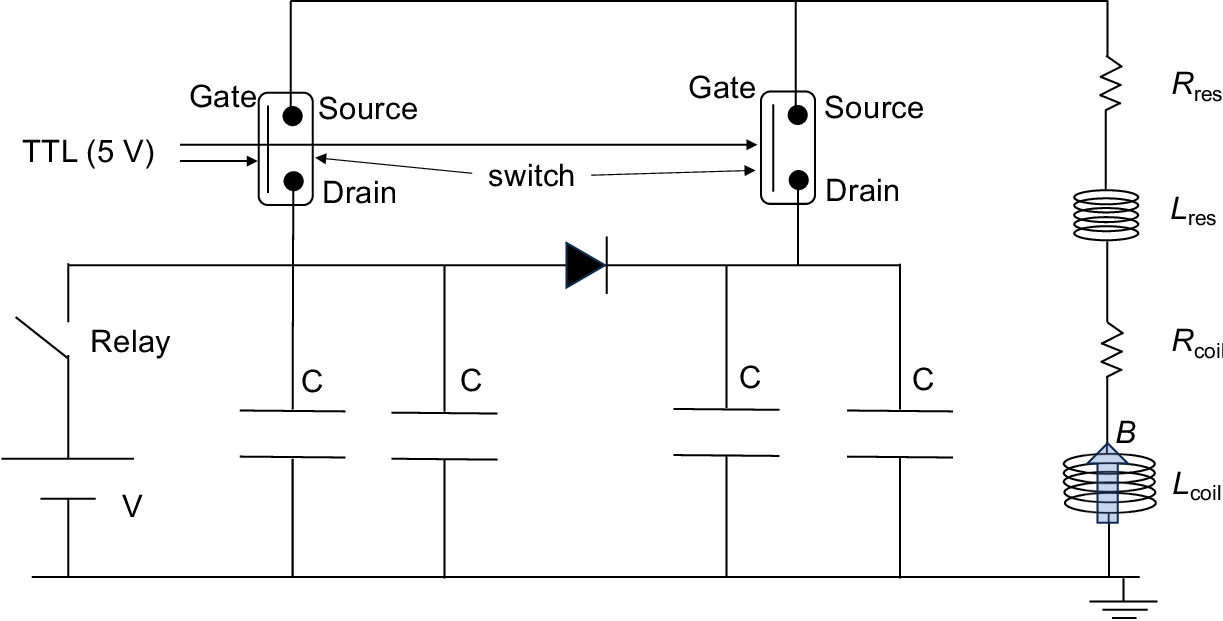}
\caption{\label{Circuit} Circuit diagram of the newly developed ultrafast pulsed magnetic field generator.}
\end{figure}

\begin{figure}[h]
\includegraphics[clip,width=0.9\columnwidth]{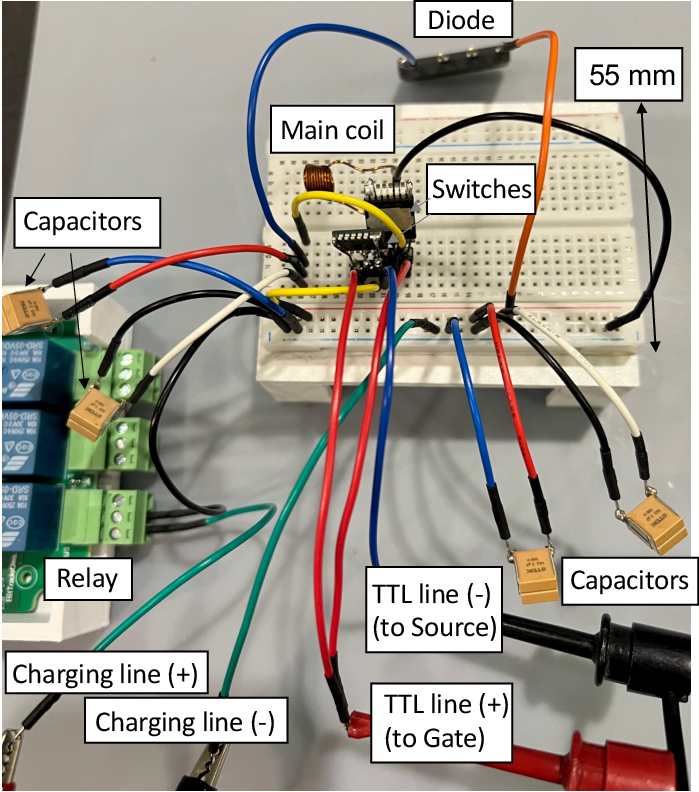}
\caption{\label{PetitBank} Picture of the new petit-bank system developed in this work.}
\end{figure}

\section{Experiments}
The circuit of our system is shown in Fig. \ref{Circuit}.
The circuit is an LCR circuit similar to that typically used in pulsed magnetic field generators.
The diode placed between two switches suppresses the interference between them, resulting in a smooth discharging waveform.
The hand-made coil was used for the main coil for generation of pulsed magnetic fields.
The inner-diameter, number of turns, and number of layers are 4.0 mm, 10 and 2, respectively.
To achieve fast discharge, the GaN high-electron-mobility-transistor (GaN-HEMT; Infineon Technologies) and a CeraLink high-power ceramic capacitor (TDK Corporation) were employed.
The residual inductances of GaN-HEMT and the CeraLink are quite small corresponding to several nH.
The GaN-HEMT switch and the CeraLink can work with the high voltage of up to 650 V and 700 V, respectively.
In addition, these devices can work in MHz-class frequency region.
The high-voltage supply (Matsusada-precision Inc.) was used to charge the capacitors.
Switching of the charging line was controlled by the relay (Bit-trade-one company). 
The magnetic field was measured by the pickup coil located at the center of the main coil.
We investigated the characteristics of the system by varying the capacitance.
Figure \ref{PetitBank} shows a picture of the new system developed in this work.
The system size corresponds to an order of 100 mm.
Small systems can be brought in other facilities, which enable us to develop the new experiment under pulsed magnetic fields such as the X-ray diffraction experiment \cite{YHMatsuda2006,AIkeda2022,AIkeda2025} under pulsed magnetic fields.
Therefore, our system also has possibility to develop the new experimental techniques under ultrafast magnetic fields.
For this initial development, the components were connected using simple cables.
As discussed later, these cables introduce residual inductances.
Replacing the cable connection with a dedicated circuit board is expected to reduce the residual inductance, leading to faster discharging.
This will be addressed in future development.
The current flowing the system was measured by the calibrated Rogowski coil.
For the demonstration of physical property measurement, the magnetization measurement of cobalt powder and Faraday rotation measurement of CdS single crystal along the $c$-axis have been performed.
The magnetization of cobalt powder was performed by the parallel-type pickup coil \cite{STakeyama2012}.
The parallel-type pickup coil consists of two coils connected in series with opposite polarities.
The d$B$/d$t$ contribution is canceled out because of the signals induced in the two coils have opposite polarities.
By inserting the sample into only one of the coils, the d$M$/d$t$ signal from the sample can be detected.
In practice, the measured signal generally contains a small residual d$B$/d$t$ contribution due to the imperfect cancellation.
The residual contribution can be removed by measuring the background signal without the sample and subtracting it from the signal measured with the sample.
For the Faraday rotation measurement, two optical fibers and a pair of linear polarizers were used with a green laser ($\lambda = 532$ nm).
The CdS single crystal was placed between two linear polarizers oriented at different angles.
The rotation of the polarization angle of light induces the change in the transmitted light intensity.
The Faraday rotation angle $\theta$ can be obtained from the measured light intensity using Malus's law.

\begin{figure}[h]
\includegraphics[clip,width=0.9\columnwidth]{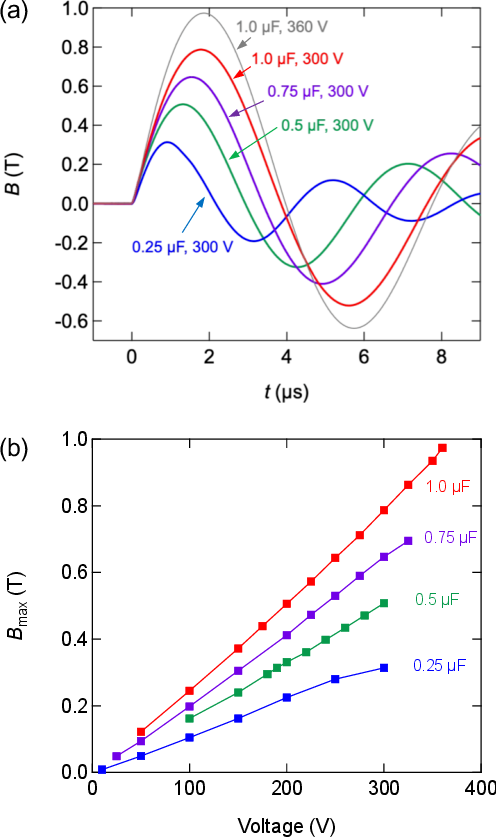}
\caption{\label{Fields} (a) Waveforms of pulsed magnetic fields at several conditions. (b) Voltage dependence of the maximum magnetic field.}
\end{figure}

\section{Results and Discussion}
Figure \ref{Fields} (a) shows the magnetic field waveforms with several conditions.
The waveform corresponds to a damped oscillation behavior which is expected to appear in this kind of circuit.
The pulse widths (the duration of the first pulse) are microsecond order in all conditions.
It also shows clear increasing behavior with increasing the capacitance.
The maximum magnetic field can reach up to approximately 1 T with the charging voltage of 360 V with the capacitance of 1.0 $\mu$F.
Figure \ref{Fields} (b) shows the voltage dependence of the maximum magnetic fields ($B_{\rm max}$) with different $C$.
The linear dependence was clearly observed.

\begin{figure}[h]
\includegraphics[clip,width=0.9\columnwidth]{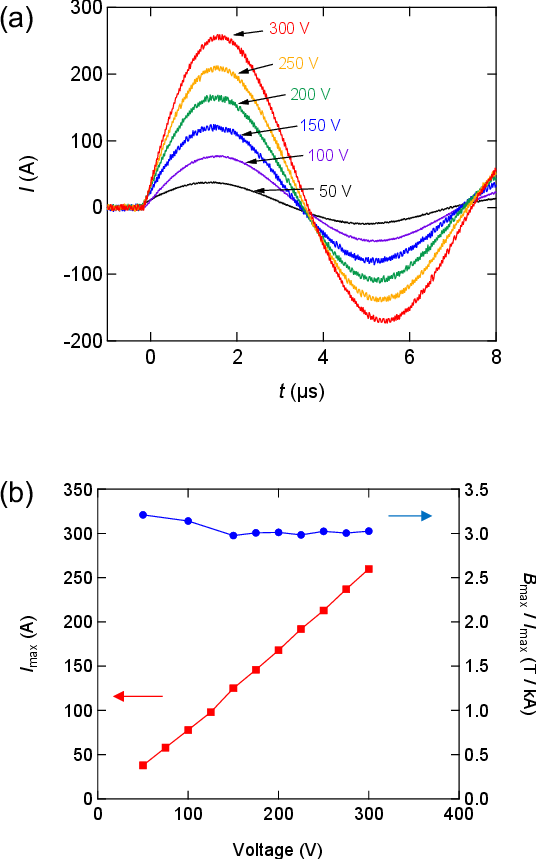}
\caption{\label{Current} (a) Waveforms of current with 1.0 $\mu$F capacitors. (b) Voltage dependence of the maximum current and the ratio between magnetic field and current ($B_{\rm max} / I_{\rm max}$).}
\end{figure}

Figure \ref{Current} (a) shows the waveforms of current with the $C$ of 1.0 $\mu$F measured by the Rogowski coil.
The observed damped oscillation behavior is the same shape of magnetic field waveforms.
The voltage dependence of the maximum current ($I_{\rm max}$) and the ratio between the maximum magnetic field to the maximum current ($B_{\rm max} / I_{\rm max}$) are shown in Fig. \ref{Current} (b).
The $I_{\rm max}$ shows linear increasing behavior with increasing the charging voltage.
The $B_{\rm max} / I_{\rm max}$ shows almost constant.
The value of $B_{\rm max} / I_{\rm max}$ as 3 (T / kA) is reasonable in this kind of system reported so far \cite{AIkeda2024Mini,AIkeda2026Rep}.

\begin{figure}[h]
\includegraphics[clip,width=0.9\columnwidth]{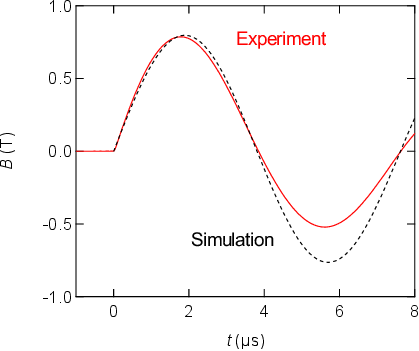}
\caption{\label{simu} The red solid and black dashed curves represent the experimental result and simulation of magnetic field, respectively.}
\end{figure}

We performed the simple calculation of time dependence of magnetic field in an LCR circuit as shown in Fig. \ref{simu}.
The simulation was carried out using the parameters corresponding to the experimental conditions, including the coil parameters as described in the Experimental section, a residual inductance of 0.4 $\mu$H, a residual resistance of 0.1 m$\Omega$, a capacitance of 1.0 $\mu$F, and a charging voltage of 300 V.
The circuit parameters used for the simulation are summarized in Table \ref{tab01}.
The $L_{\rm res}$ and $R_{\rm res}$ are mainly attributed to the connecting cables.
The simulation result shows  quantitative good agreement with the experimental result.
The damping of the magnetic field in experiment is faster than in the simulation.
This discrepancy is probably attributed to the characteristics of the GaN-HEMT switches and CeraLink capacitors, since these advanced powder devices can exhibit complex voltage- and frequency-dependent behavior.
Nevertheless, the first pulse is well reproduced by the simulation, suggesting that these devices operate as expected, at least during the first pulse.

\begin{table}[h]
\centering
\caption{Parameters for the simulation. $L_{\rm coil}$ ($\mu$H), $L_{\rm res}$ ($\mu$H), $R_{\rm coil}$ (m$\Omega$), $R_{\rm res}$ (m$\Omega$), and $C$ ($\mu$F) represent the inductance of coil, residual inductance, resistance of coil, residual resistance, and capacitance, respectively.}
%\resizebox{0.95\columnwidth}{!}{
\normalsize
\begin{tabular}{cccccc} \hline \hline
$L_{\rm coil}$ ($\mu$H) & $L_{\rm res}$ ($\mu$H)  & $R_{\rm coil}$ (m$\Omega$) & $R_{\rm res}$ (m$\Omega$) & $C$ ($\mu$F) &\\ \hline
1.1 & 0.4 & 34 & 0.1 & 1.0  \\
\hline \hline
\end{tabular}
%}
\label{tab01}
\end{table}

\begin{comment}
\begin{table}[h]
\centering
\caption{Maximum magnetic field ($B_{\rm max}$), maximum current ($I_{\rm max}$), and pulse width in the experiment and the simulation with $C = 1.0 \mu$F and $V$ = 300 V.}
%\resizebox{0.95\columnwidth}{!}{
\normalsize
\begin{tabular}{cccccc} \hline \hline
 & $B_{\rm max}$ (T) & $I_{\rm max}$ (A) & pulse width ($\mu$s)\\ \hline
experiment \quad & 0.8 \quad & 250 \quad & 3.8 \quad  \\ \hline
simulation \quad & 0.8 \quad & 242 \quad & 3.8 \quad \\
\hline \hline
\end{tabular}
%}
\label{tab02}
\end{table}

The maximum magnetic field ($B_{\rm max}$), the maximum current ($I_{\rm max}$), and the pulse width obtained from both the experiment and the simulation are summarized in Table \ref{tab02}.

\end{comment}

\begin{figure}[h]
\includegraphics[clip,width=0.9\columnwidth]{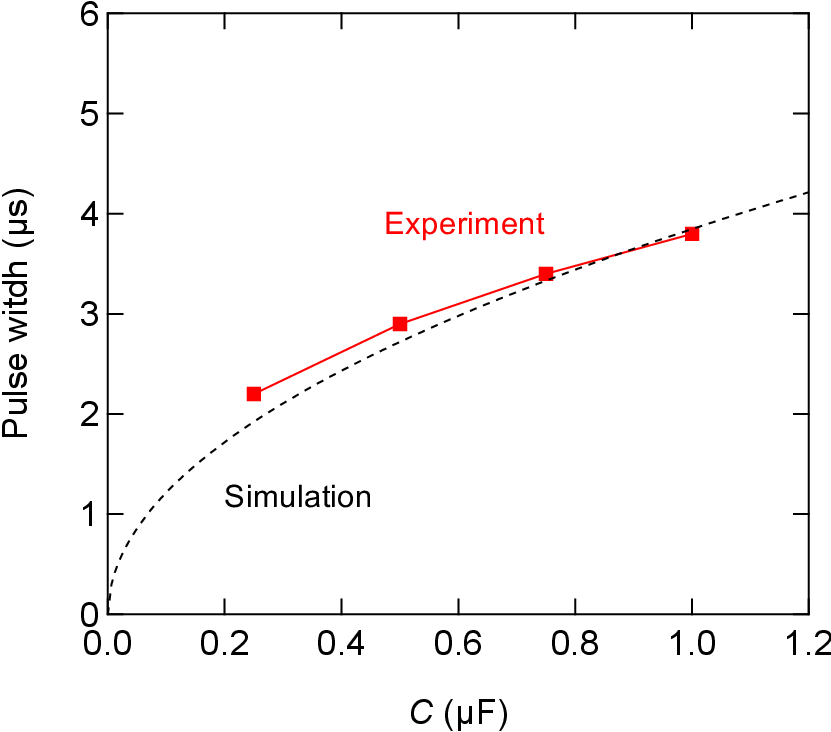}
\caption{\label{Width} Red curve shows the capacitance dependence of the pulse width. Black dashed line shows the calculation result of pulse width based on the simple LCR circuit.}
\end{figure}

Figure \ref{Width} shows the dependence of the pulse width on the capacitance $C$.
This dependence is described by the expression

\begin{equation}
T/2 = \pi / \sqrt{\frac{1}{(L+L_{\rm res})C} - \frac{(R+R_{\rm res})^2}{4(L+L_{\rm res})^2}}
 \end{equation}

where $T$ is period of the damped oscillation ($T/2$ corresponds to the pulse duration of the pulsed magnetic field). The parameters in Table \ref{tab01} are also used for this simulation.
The simulation result shows the good quantitative agreement.

The total energy stored in the capacitor with the charging voltage of 300 V is

\begin{equation}
E = CV^2/2 = 1.0 \times 10^{-6} \times 300^2 / 2 = 45 {\rm [mJ]}
\end{equation}

Typical pulsed magnetic field generator requires the large energy corresponding to 10$^2$ to 10$^5$ [kJ] to generate the pulsed magnetic field of 10$^1$ or 10$^2$ T.
Regarding the energy efficiency [Tesla / J], the new system has large efficiency corresponding to approximately 20 [Tesla / J] whereas it is less than 1.0 in the typical pulsed field generators.

\begin{figure}[h]
\includegraphics[clip,width=0.9\columnwidth]{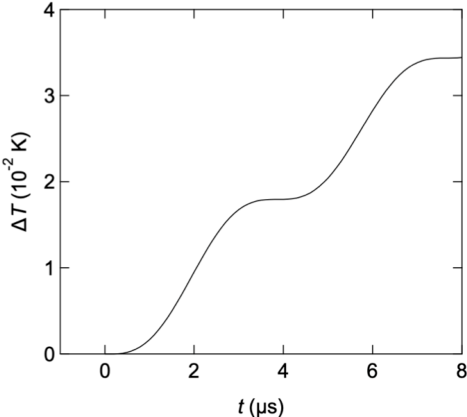}
\caption{\label{simu_temp} Estimation of the temperature increase of the main coil coming from the Joule heating.}
\end{figure}

Figure \ref{simu_temp} shows the estimation of temperature increase ($\Delta T$) of the main coil.
The $\Delta T$ is below 0.02 K by a single pulse, which indicates that the $\Delta T$ can be negligible. 
It worth noting that the coil can be placed directly in a low-temperature environment, providing greater flexibility for experiments under low-temperature conditions.

\begin{figure}[h]
\includegraphics[clip,width=0.9\columnwidth]{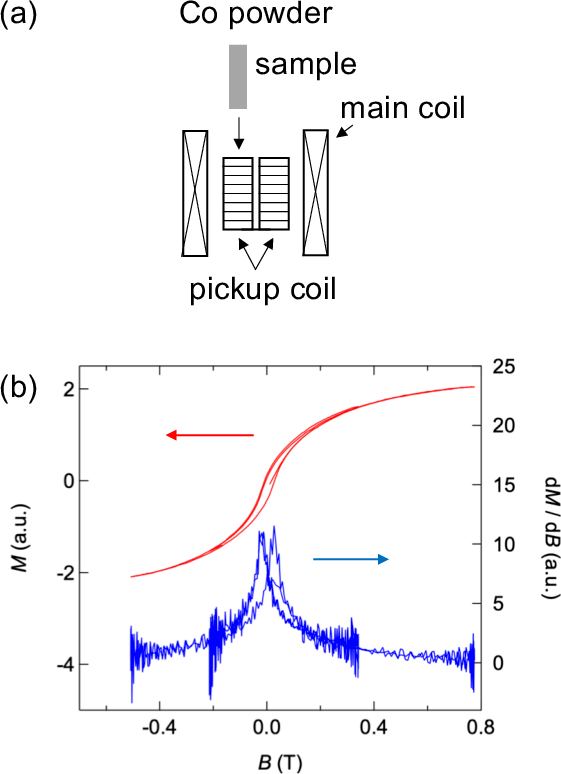}
\caption{\label{M_Co} (a) Schematic of experimental setup of magnetization measurement on cobalt powder sample. (b) Red and blue curves represent magnetization curve and its field derivative, respectively.}
\end{figure}

Finally, we present the magnetization measurement of cobalt powder and the Faraday rotation measurement of CdS single crystal as demonstrations of the applicability of the developed system to physical property measurements.
Figure \ref{M_Co} (a) shows schematic of experimental setup of magnetization measurement on cobalt powder sample.
The magnetization curve (red) and its field derivative (blue) of cobalt powder are shown in Fig. \ref{M_Co} (b).
These data were obtained from a single pulse experiment.
The data shows clear hysteresis  associated with the ferromagnetic nature of cobalt.
The coercive field was estimated to be 0.025(5) T, which is consistent with previously reported values \cite{HLuo2005}.

\begin{figure}[h]
\includegraphics[clip,width=0.9\columnwidth]{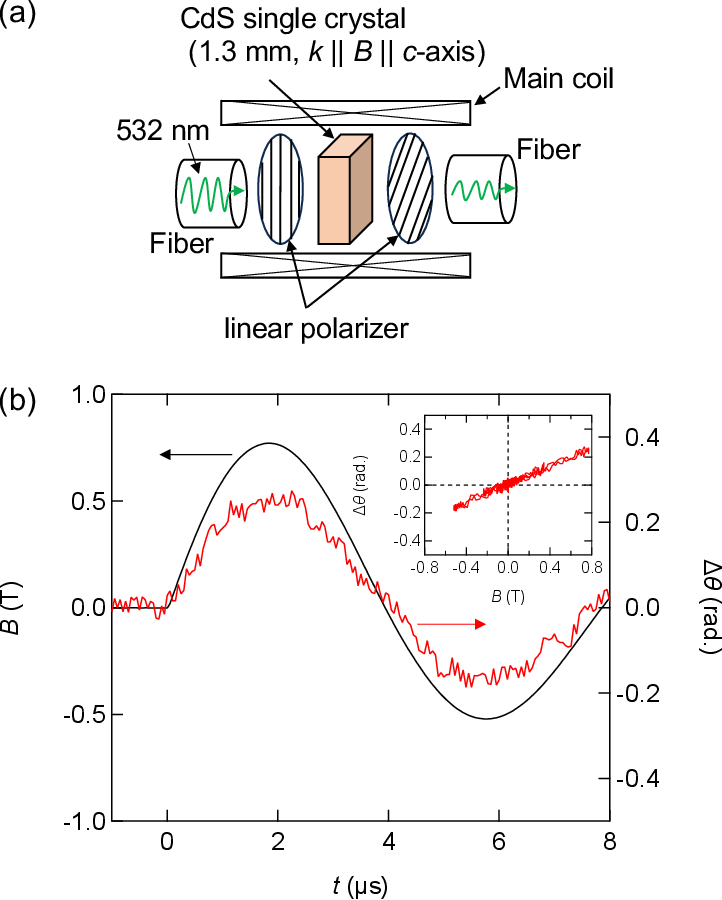}
\caption{\label{FaradayCdS} (a) Schematic of experimental setup of Faraday rotation measurement on CdS single crystal. (b) Faraday rotation ($\Delta \theta$) of CdS single crystal (red) and magnetic field (black) as functions of time. Inset shows the magnetic field dependence of $\Delta \theta$.}
\end{figure}

A schematic of experimental setup of Faraday rotation measurement of CdS single crystal is shown in Fig. \ref{FaradayCdS} (a).
Figure \ref{FaradayCdS} (b) shows the result of Faraday rotation measurement of CdS single crystal.
The black and red curves show the magnetic field and $\Delta \theta$ as functions of time, respectively.
The inset shows the $\Delta \theta$ as a function of magnetic field.
From this result, the Verdet constant has been estimated to be 14.5 (deg. T$^{-1}$ mm$^{-1}$).
This value is consistent with the values reported elsewhere \cite{VVDruzhinin1995,YImanaka2004}.

\section{Conclusion}
We have developed a new compact nondestructive generator for microsecond-class ultrafast pulsed magnetic fields, referred to the "petit-bank system", using commercially available ultrafast power devices.
The newly developed system can generate the pulsed magnetic field of up to approximately 1 T with pulse duration of several microseconds.
In this paper, we have presented the design and performance of the system.
The properties of the system can be described by a simple simulation taking into account the residual inductance and resistance.
To the best of our knowledge, this is the first demonstration of a compact system capable of generating microsecond-class ultrafast pulsed magnetic fields.
The key to achieving such ultrafast pulsed magnetic fields is the use of state-of-the-art ultrafast power devices such as GaN-HEMT and CeraLink capacitors.
In addition, we have demonstrated physical property measurements using the petit-bank system, confirming its applicability to condensed matter experiments.

The petit-bank system also offers considerable potential for further development and applications.
Its high extensibility is one of the key advantages of newly developed system.
By increasing the number of capacitors, higher magnetic fields can be generated.
Based on a simple estimate, the magnetic field is expected to reach approximately 10 T by increasing the capacitance and the charging voltage to 50 $\mu$F and 600 V, respectively.
Magnetic fields of up to 10 T will expand the range of phenomena that can be investigated, including those in which the magnetic field sweep rate plays a crucial role, such as spin crossover and quantum tunneling. 
Another advantage is the absence of the initial noise originating from the air-gap switches used in destructive pulsed magnets \cite{FHerlach1973, OPortugall1997,NMiura2001}.
The petit-bank system therefore enables precise physical property measurements under ultrafast magnetic fields.

\section{Acknowledgement}
We thank H. Sawabe, K. Sudo, and K. Noda for fruitful discussions, This research was supported by JSPJ KAKENHI Grants Numbers 23H01117, JP23H04859, JP23H04860, JP23H04861, 24K17003, and JP25H02118.

\section{Data Availability}
The data that support the findings of this study are available from the corresponding author upon reasonable request..

\begin{comment}
\appendix
\end{comment}

\nocite{*}
\bibliography{PetitBank}% Produces the bibliography via BibTeX.

\end{document}